# The contribution of 211 particles to the mechanical reinforcement mechanism of 123 superconducting single domains

J.-P. Mathieu[1], I.G. Cano[1], T. Koutzarova[1], A. Rulmont[1], Ph. Vanderbemden[2], D. Dew-Hughes[3], M. Ausloos[4], R. Cloots[1,*]

[1] S.U.P.R.A.T.E.C.S., Chemistry Institute B6, University of Liege, Sart Tilman, B-4000 Liege, Belgium
[2] S.U.P.R.A.T.E.C.S., Montefiore Electricity Institute B28, University of Liege, Sart Tilman, B-4000 Liege, Belgium
[3] Department of Engineering Science, Oxford University, Parks Road, Oxford OX1 3PJ, UK
[4] S.U.P.R.A.T.E.C.S., Physics Institute B5, University of Liege, Sart Tilman, B-4000 Liege, Belgium

* Corresponding author. Tel.: +32-4-366 34 36; fax: +32-4-366 34 13.
E-mail address: rcloots@ulg.ac.be (R. Cloots).

ABSTRACT:

Hardness and fracture toughness of Dy-123 single-domains were studied by Vickers micro-indentation. A significant anisotropy of the mechanical properties was observed. Hardness tests give higher values when performed in (001) planes rather than in planes parallel to the *c*-axis. Moreover cracks pattern around the indentation follows preferential orientation in planes parallel to the *c*-axis whereas a classical four-cracks pattern is observed in the (001) planes. It has been possible to show the crucial role played by the 211-particles in the deviating mechanism of cracks and the relevance of the 211-particle distribution high homogeneity in the material.

## I. INTRODUCTION

Since the discovery of superconductivity in the YBCO system, considerable progress has been made in preparing samples with high critical currents and strong pinning forces. The good superconducting properties of $YBa_2Cu_3O_{7-\delta}$ (Y-123) single-domains open a wide field of potential applications, *e.g.* in magnetic bearings [1,2] or as trapped flux magnets [3,4].

Nowadays, the top-seeded melt-textured technique is considered to be the best process to obtain large-grain samples in which the crystal orientation is controlled through a seeding technique [5-7]. Single-crystals of $SmBa_2Cu_3O_{7-\delta}$ (Sm-123) or $NdBa_2Cu_3O_{7-\delta}$ (Nd-123) serve as effective seed for Y-123 due to the similar lattice parameters of these 123-compounds. Typically the seed is placed on the top of the sample, promoting epitaxial growth of Y-123 at the Nd-123 — Liquid+$Y_2BaCuO_5$ (Y-211) interface [8]. Moreover, a thermal gradient can be applied in order to improve the domain growth. Y-Ba-Cu-O quasi single-crystals up to 70 mm have been processed by this method [9].

Recently it has been demonstrated that $DyBa_2Cu_3O_{7-x}$ single-domains (Dy-123), prepared by the same route [10], can achieve better superconducting properties, *i.e.* higher $J_c$ and $H_c$, than Y-123 [10-13]. Therefore, Dy-Ba-Cu-O system is one of the most promising candidates for high field engineering applications and is studied hereafter.

However, as shown by Ren *et al*. [14], interactions between trapped field and critical current induce important outward pressure that can cause the cracking and the fracture of the material. In practical applications, the superconductor will be subjected to many temperature cycles, with large temperature variations between room and liquid nitrogen temperatures. These temperature changes are expected to promote thermal fatigue and, hence, to cause premature failure of the material. For these reasons, it can be easily understood that mechanical properties will limit severely the high field engineering applications. Therefore, the improvement of mechanical properties is a significant problem for the achievement of large trapped fields.

Numerous studies on YBCO bulk materials focus on the optimisation of their microstructure and electrical properties [5,8,10,13,15-18], but little is known about the mechanical properties of these materials.

Superconducting Dy-123 single-domain materials are characterized by a complex microstructure, consisting of a dispersion of 211 particles, in a 123-matrix, both having different thermal expansion coefficients, which can be an intrinsic source of micro-cracks [19]. Different thermal stresses arising in the 123-phase during cooling have been proposed as the origin of cracking, and a critical 211-particle size for the nucleation of a-b microcracking has been estimated [19].

From considerations based again on the thermal and elastic mismatch effects between 211 and 123 phases, and from microstructural data, Goyal and co-workers [20] have argued that the 211 particles can be considered as a reinforcement material contributing to the enhancement of the fracture resistance behaviour of 123 matrix through energy dissipation due to interfacial delamination and crack bridging. The former mechanism of energy dissipation may result from the strong chemical bonding existing between both phases when a high modulus reinforcement particle is added to a lower modulus matrix [21]. Measurements made by an ultra-low load indentation technique indicate that 211 particles are characterized by a higher average Young s modulus (213 GPa) than the 123 matrix (180GPa) [20].

To understand and to control the mechanical properties of 123/211 superconducting single domains, it is necessary to take into account the strong anisotropy of the 123 matrix. Tancret *et al.* [22] propose a description of the microstructure as a stacking of 123 layers cracked along the (001) crystallographic planes with inserted 211 particles joining them. Crack deviation mechanism by the 211 particles has been observed and the presence of 211 particles has been considered as a key factor for enhancing the toughness of 123 melt-textured materials for only (100) and (010) propagating cracks. Thermal shock resistance has been also improved by the presence of 211 particles up to a certain degree of 211 concentration, the 211 particles acting as solid links between 123 cracked layers. Tancret *et al.* [22] also reported that microcracks of the 123-matrix along the (001)-planes are created due to the difference in thermal expansion coefficients of the 123 and 211 phases. The microcracks are more pronounced when the concentration of 211 particles increases, yielding an increase of the thermal stresses during cooling, and thus a decrease in thermal shock resistance.

From general considerations, several mechanisms have been proposed in the literature [23] for increasing fracture toughness in ceramic composites:

(1) A local concentration of micro-cracks, by dispersing particles that can produce a higher density of micro-cracks than the matrix itself, results in an increase of the toughness of the material. Notice also that the presence of micro-cracks can induce a decrease of the thermal shock resistance as mentioned before for 123 materials, by promoting crack propagation during cooling. The positive effect of micro-cracking for increasing fracture toughness in a melt-textured 123 material is thus very controversial.

(2) The presence of particles acting as barriers to cracks propagation, dispersed in the matrix, that result in at least temporary pinning of the crack front, results also in an increase of the toughness of the material. This effect can also be putted into evidence in the case of 123 melt-textured materials, taking into account the stress fields developed around the 211 particles [22].

(3) Finally, cracks wake bridging has become a widely cited mechanism of toughening based on observation of particles bridging the wake zone of cracks. Such a wake bridging mechanism of toughening was also reported for melt-textured materials, taking into account the strong anisotropy of their microstructure [20]. The wake bridging mechanism of toughening increases as particle size increases whereas pinning increases with decreasing particle size.

Each mechanism can contribute to a variable extent to the reinforcement of the mechanical properties of 123 melt-textured materials, due to their complex microstructure, related to the presence of 211 particles dispersed in the 123 matrix.

In this paper, the hardness and fracture toughness of Dy-Ba-Cu-O superconducting single-domains is deduced from micro-indentation measurements. Attention is focussed on the important role played by the 211-particles on crack propagation, taking into account the difference in both Young s moduli and thermal expansion coefficients of 211 and 123 phases. Mechanisms of toughening are discussed based on the microstructural data compiled for these materials.

## II. EXPERIMENTAL PROCEDURE

$DyBa_2Cu_3O_{7-x}$ (Dy-123) and $Dy_2BaCuO_5$ (Dy-211) powders, used as precursors to prepare materials for mechanical properties evaluation, have been produced by the solid-state synthesis route from $Y_2O_3$, $BaCO_3$ and CuO powders. Dy-123 and Dy-211 were mixed in 79:20 weight ratio with 1 wt% of $CeO_2$ in order to avoid the Ostwald ripening of the Dy-211 particles in the liquid phase during the melt process [24,25]. The powder mixture was pressed uniaxially (15T) to form parallelepipedic and cylindrical pellets (20×20×10mm and 10mm×15mm in diameter, respectively).

Pellets were melt-textured under air using a Nd-123 single-crystal seed, prepared by the flux method [26], placed on the top of the pressed powder [5,8]. The growth starts from the seed and generally proceeds throughout the entire volume of the precursor material. In this case, a large quasi single-crystal can be obtained. The thermal cycle applied is given in figure 1(a). After the growth process, samples were oxygenated by applying the thermal cycle given in figure 1(b) under flowing $O_2$.

The mechanical properties (hardness and fracture toughness) have been evaluated through Vickers micro-indentation measurements. Pellets were polished using SiC abrasives followed by diamond pastes. Indentations were performed with a pyramidal diamond indenter. Two different indentation loads were used: 500g and 1000 g.

The crystal axis of as-obtained samples were determined by optical polarized light microscopy (Olympus VANOX AHMT3), and confirmed by X-ray diffraction analysis (Siemens D-5000). Crack propagation and the presence of toughening induced by 211 particles were also observed by scanning electronic microscopy (Philips ESEM XL30 FEG).

## III. RESULTS AND DISCUSSION

In order to evaluate the mechanical behaviour, micro-hardness measurements were performed in different zones either parallel or perpendicular to the *c*-axis. Micro-hardness values ($H_V$) were been estimated according to equation 1:

$$H_V = \frac{1854,4 \, ?F}{d?} \quad (GPa) \qquad (eq.\ 1)$$

where $F$ is the contact load (expressed in N) and $d$ is the length of the diagonal (in m) as shown in figure 2.

Toughness ($K_c$) can be evaluated, when the cracks pattern is radial and well-defined, according to equation 2:

$$K_c = \xi \sqrt{\frac{E}{H_V}} \frac{F}{c_0^{3/2}} \quad (Pa\ m^{-1/2}) \qquad (eq.\ 2)$$

$\xi$ is a material-independent constant for Vickers-produced radial cracks determined by Antis G.R. *et al.* [27] and equal to $0,016 \pm 0,004$. $E$ is the Young modulus (in GPa) and $c_0$ is the average cracks length (in m) as shown in figure 2.

More than ten indentations were performed at different locations in planes parallel and perpendicular to the *c*-axis. Depending on the crystallographic orientation of the single grain, different behaviours have been observed. Results are summarised in Table 1.

In planes perpendicular to the *c*-axis, *i.e.* (001) or *a/b* planes, the average hardness (6.68 GPa) value is in good agreement with that obtained by other groups for the Y-Ba-Cu-O system [28-30]. This result lies between the Y-123 single-crystal hardness value in the same planes (8.7GPa) [31] and the hardness value reported for Y-123 polycrystalline samples (5GPa) [32]. The crack pattern induced by the indentation is characteristic of typical four-cracks propagation behaviour in an isotropic medium where it is clearly observed that the cracks are perpendicular and similar in length (see figure 3-a). In these conditions, it is possible to evaluate the toughness of the material according to equation (2), if the Young s modulus of the system is known. Goyal A. *et al.* [20] have determined the Young s modulus to be 143 ± 4 GPa and 182 ± 4 GPa for (001) planes and planes parallel to the *c*-axis respectively. If it is assumed that the primary wedge opening strains associated with the indentation process are perpendicular to the direction of the cracks, the Young s modulus value in the direction perpendicular to the crack face has to be employed, *i.e.* 182 GPa, in order to determine the fracture toughness for indentation on (001) planes.

In planes parallel to the *c*-axis, the average hardness value is found to be lower than that reported for (001) planes (see Table 1). Moreover, the cracks have been propagating perpendicularly to the *c*-axis following the micro-cracks already present in the material whatever the initial orientation of the indenter, *i.e.* there is a preferential orientation for cracks to propagate (see figure 3-b). This behaviour is consistent with an easy path along the *a/b* planes for crack propagation through the sample. However, it is not possible in this case to evaluate correctly the fracture toughness.

Dy-Ba-Cu-O presents thus a high anisotropy in its mechanical behaviour, which is strongly dependent on the crystallographic orientation of the system.

In addition, SEM analyses have been performed in order to more precisely determine the crack propagation mechanism(s) in these materials. As can be seen in figure 4-*a*, the crack deviates from its initial path, when it reaches a 211-particle (in white in the micrographs), and tends to propagate around the 211 particle, due to the residual mismatch stress field around them, which is not necessarily the case when micro-cracks are induced by thermal stress during cooling [19]. This latter fact can be explained by considering the difference in thermal expansion coefficients between 211 and 123, parallel to the c-axis. When cracks are created by indentation, the key parameter to take into account is the Young s modulus. The Young s modulus of 211-particles is larger than the one of 123-matrix. It has been kept in mind that the incorporation of high modulus particles in a low-modulus matrix results in a reduction of the tensile stress at the tip of a crack running through the matrix [20]. It is clear from these observations that the 211-particles play a first significant role in the Dy-123 toughening mechanism by deviating the cracks. Another possible mechanism of toughening is suggested in these materials, and is illustrated in figure 4-*b* and *c*. A wake bridging zone is essentially observed for an area with a high concentration of 211 particles. 211 particles play thus another retardant role in the crack propagation mechanism by crack bridging, resulting in an increase of the fracture toughness of the material. Cracks thus propagate inhomogeneously in these materials depending on the distribution of the 211 particles.

In order to illustrate this effect, optical micrographs of indentations performed again in a plane parallel to the *c*-axis are shown in figure 5-*a* and *b*. It is readily visible that the 211-particle distribution is not homogeneous: more cracks are found in the area of the sample where there are few 211-inclusions. This is once again an experimental proof of the role played by the 211-particles in the toughening mechanism and this illustrates the necessity of controlling the homogeneity of these materials. Consequently it seems very important to control the 211 distribution in order to enhance the mechanical properties.

## IV. CONCLUSIONS

Mechanical properties of Dy-123 materials, *i.e.* hardness and toughness, have been measured through Vickers indentation. Average hardness values are similar to those observed for Y-Ba-Cu-O system. A significant anisotropy of the mechanical behaviour is observed. It is shown that hardness is higher when the measurement is performed in (001) planes than in planes parallel to the *c*-axis. The strongest difference concerns the crack propagation pattern versus the crystallographic orientations. Contrary to the behaviour found in the (001) planes where a classical four-cracks pattern is observed, a preferential crack propagation is shown in planes parallel to the *c*-axis.

The crucial role played by the 211-particles on the toughening mechanism has been also demonstrated. It has been shown that cracks are deviated by 211-particles, and that a wake crack bridging zone appears when a high concentration of 211 particles is present. Mechanical properties in these materials are strongly dependent on the distribution of 211 particles in the bulk. Thermal stresses induce micro-cracks at the interface between 123 and 211 phases due to the difference in thermal expansion coefficients, whereas macro-cracks induced by indentation are limited by the difference in Young s modulus between 123 and 211 through different mechanisms of material toughening. In each case, homogeneity of the 211-particles distribution is a crucial point to achieve good mechanical properties.

The RE-Ba-Cu-O single-domains are still rather brittle materials and the poor mechanical behaviour limits engineering applications involving high magnetic fields. Nevertheless, it has been shown that mechanical properties can be improved by controlling the microstructure, *i.e.* the 211-particles distribution. In fact, this is important not only for mechanical properties, but also for superconducting properties since 211-particles are good flux-pinning centres.

## V. ACKNOWLEDGEMENTS

The materials preparation and the characterizations are part of the doctorate thesis of J.-P. Mathieu who thanks FRIA (Fonds pour la Formation la Recherche dans l'Industrie et dans l'Agriculture), Brussels for financial support. Ms. Koutzarova and Ms. Garcia have been financially supported by the European Supermachines Research Training Network (HPRN-CT-2000-0036).

# VII. TABLE AND FIGURES

## Table 1

|  | Hardness (GPa) | Toughness (MPa m⁻) |
|---|---|---|
| ⊥ *c*-axis planes | 6.68 ± 1.09 | 0.96 ± 0.17 |
| // *c*-axis planes | 4.96 ± 0.66 | / |

**Table 1:** Average hardness and toughness of Dy-Ba-Cu-O single-domains versus the planes on which the measurement was performed

## Figure 1

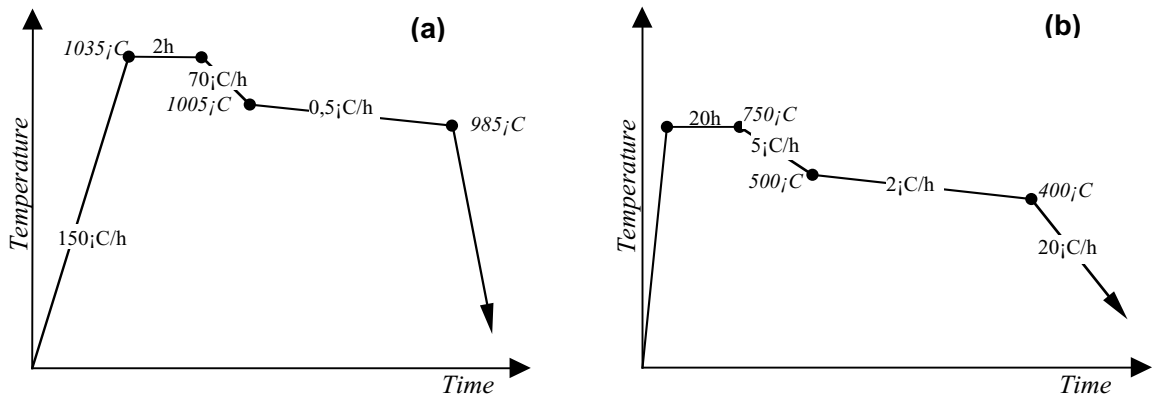

**Figure 1:** Thermal cycle for the fabrication of Dy-Ba-Cu-O bulk samples: (a) melt-texturing; (b) oxygenation

**Figure 2**

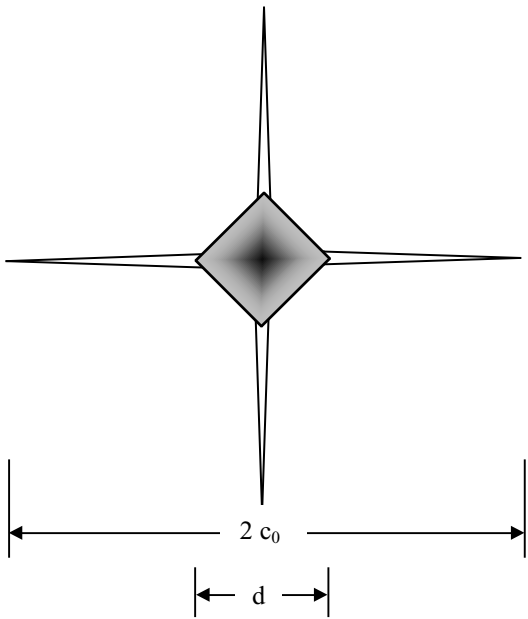

**Figure 2:** Schematic drawing of Vickers-produced indentation-fracture printing showing dimensions of the crack, $c_0$, and the diagonal of the indentation, $d$.

**Figure 3**

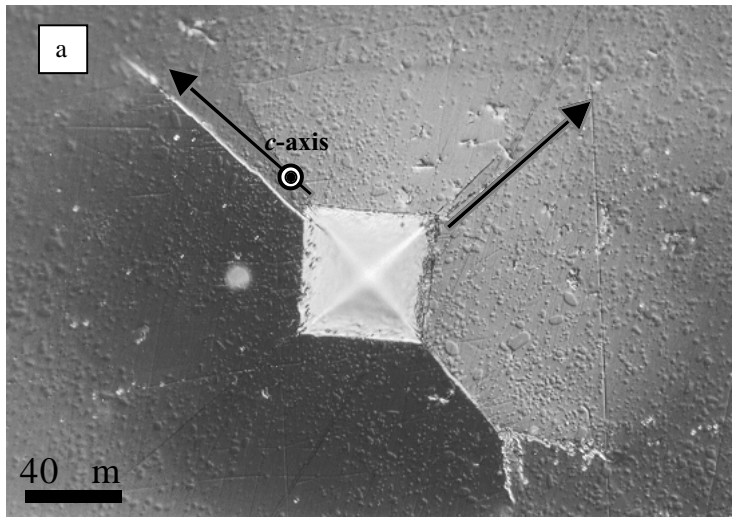

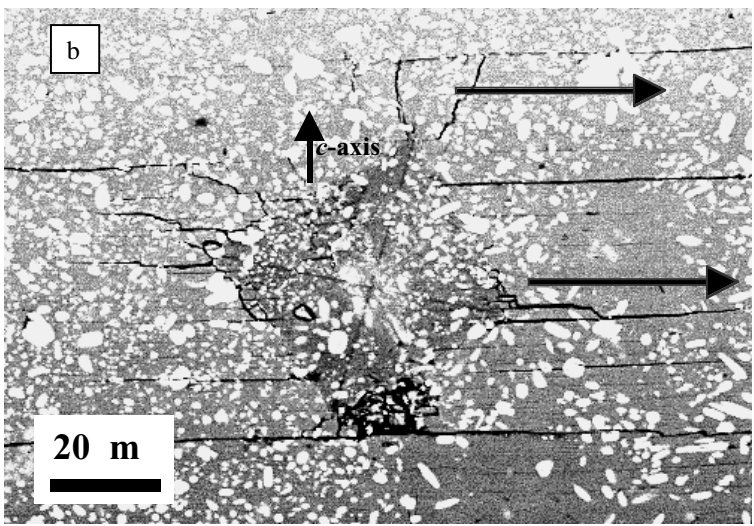

**Figure 3:** (a) Optical micrograph of an indentation performed in a (001) plane, radial crack pattern can be observed; (b) SEM micrograph of an indentation in a plane parallel to *c*-axis, cracks propagate following a preferential orientation along the *a/b* planes.

**Figure 4**

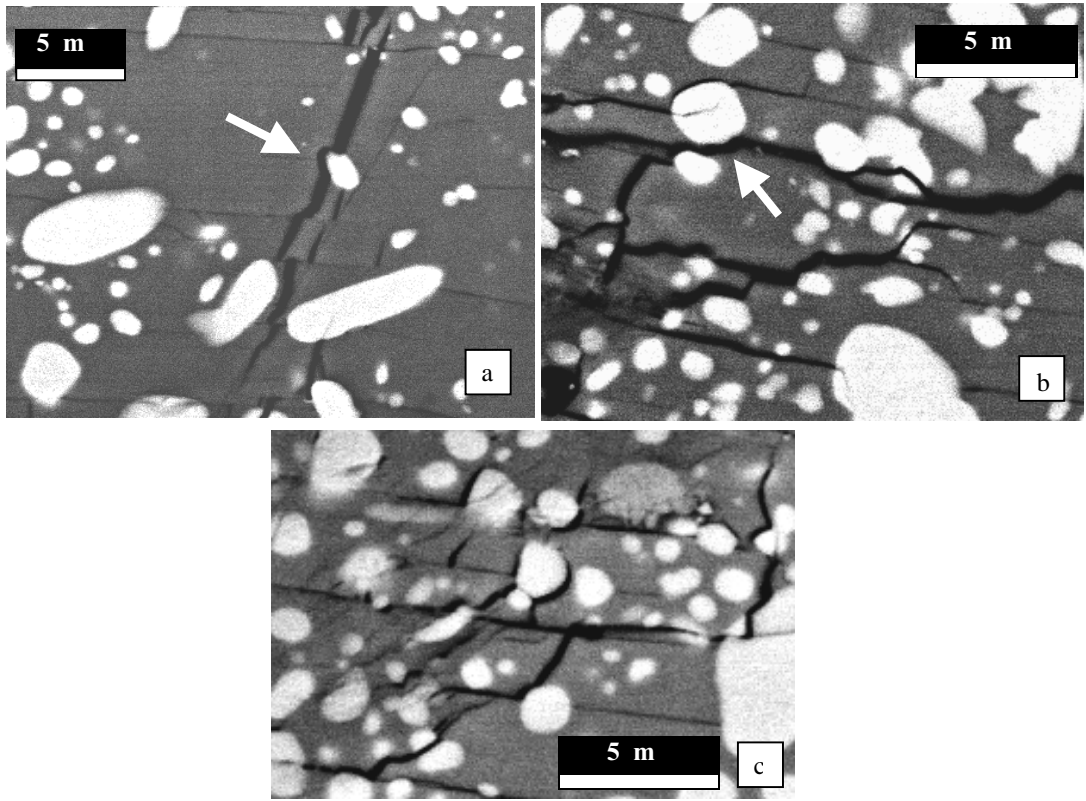

**Figure 4:** Crack propagation in the 123-matrix: (a) Crack path modified by 211-particles; (b) and (c) crack jump from one 211-inclusion to another due to the high proximity between 211-particles.

**Figure 5**

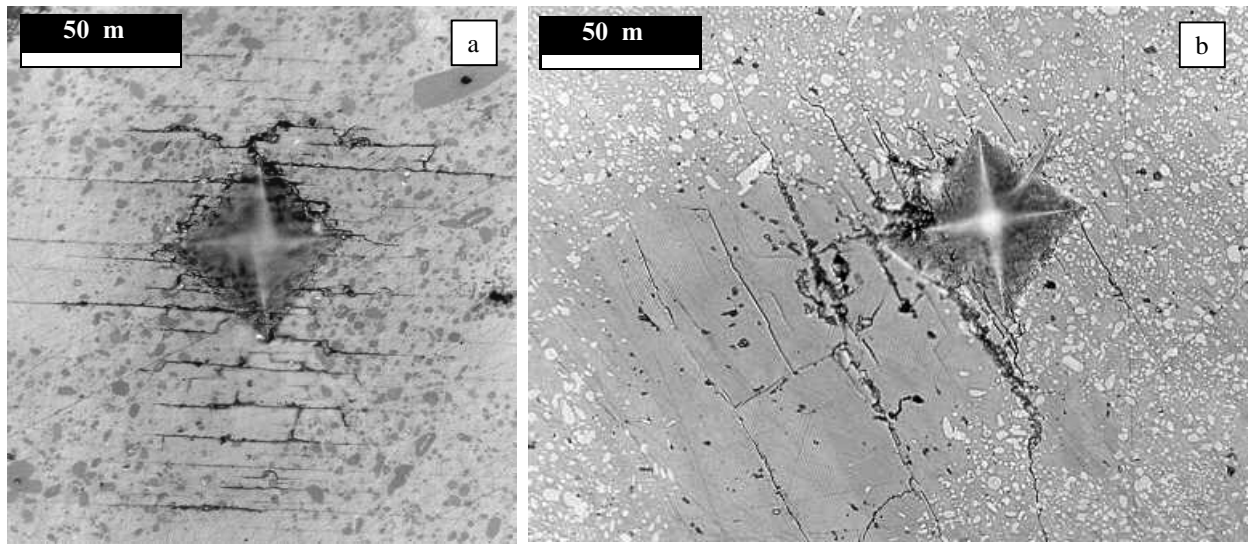

**Figure 5-a and b:** Optical micrographs of indentations performed in a plane perpendicular to the *c*-axis. It can be seen that the number of cracks is higher in the low 211-particles content area.